\documentstyle[sprocl]{article}

\def\be{\begin{equation}}

\def\ee{\end{equation}}

\def\bea{\begin{eqnarray}}

\def\eea{\end{eqnarray}}

\begin{document}

\title{PHYSICS, COSMOLOGY AND EXPERIMENTAL SIGNATURES
OF A POSSIBLE NEW CLASS OF SUPERLUMINAL PARTICLES}

\author{ L. GONZALEZ-MESTRES}

\address{Laboratoire de Physique Corpusculaire, Coll\`ege
de France,\\ 11 place Marcellin-Berthelot, 75231 Paris Cedex 05 , France
\\ and \\ Laboratoire d'Annecy-le-Vieux de 
Physique des Particules, B.P. 110 \\ 74941 Annecy-le-Vieux Cedex, France}




\maketitle\abstracts{
The apparent Lorentz invariance of the laws of physics does 
not imply that space-time is indeed minkowskian.
We consider a scenario where Lorentz invariance is 
only an approximate property of equations
describing a sector of matter at a given scale and 
superluminal sectors of matter exist related to new degrees of freedom
not yet discovered experimentally. The new particles would 
not be tachyons: they may feel different minkowskian space-times
with critical speeds much higher than $c$ (speed of light)
and behave kinematically like ordinary particles apart from the difference in
critical speed. Superluminal particles may 
provide most of the matter at cosmic scale, and be mainly dark.
We present a discussion of possible theoretical, 
cosmological and experimental consequences of such a scenario,
with particular emphasis on problems related to 
the identification of dark matter.
}
  
\section{Lorentz invariance and superluminal sectors}

\subsection{Relativity and sine-Gordon solitons}

In textbook special relativity, minkowskian geometry is an intrinsic property
of space and time. However, a look to various dynamical systems
would suggest a more flexible approach to the relation between matter and
space-time. Lorentz invariance can be viewed as a symmetry of the equations
of motion, in which case no reference to absolute properties of space and time
is required. In a two-dimensional galilean space-time, the equation:
\equation
\alpha ~\partial ^2\phi /\partial t^2~-~\partial ^2\phi /\partial x^2~=~F(\phi )
\endequation
with $\alpha ~=~1/c_o^2 $ and $c_o~=$ critical speed, remains unchanged under
"Lorentz transformations" leaving invariant the squared interval $ds^2
~=~dx^2~-~c_o^2dt^2$ , so that matter made with solutions of equation (1) would
feel a relativistic space-time even if the real space-time is actually galilean
and if an absolute frame exists in the underlying dynamics beyond the wave
equation. A well-known example is provided by the solitons of the sine-Gordon
equation, obtained taking in (1): $F(\phi )~=~(\omega /c_o)^2~sin \phi $ . 
A two-dimensional universe made of sine-Gordon solitons plunged in a galilean
world would feel a two-dimensional minkowskian space-time 
with the laws of special
relativity. Information on any absolute rest frame would be lost by the
solitons.

1-soliton solutions of the sine-Gordon equation are known to exhibit
"re- lativistic" particle properties, e.g. $E$ ( energy) =
$E_o~(1~-~v^2/c_o^2)^{-1/2}$ , where  $v$ is the soliton speed and $E_o$ its 
rest energy,
so that everything looks perfectly
"minkowskian" even if the basic equation derives from a galilean world 
with an absolute rest frame. Similarly, in the real world, the speed of light
$c$ could be just the sectorial critical speed of a part matter (the "ordinary"
particles), instead of a universal critical speed deriving from absolute
geometric properties of space and time as usually stated in relativity theory. 

\subsection{Superluminal particles}

If Lorentz invariance is only an approximate property of equations describing a
sector of matter above a given distance scale, and absolute frame (the "vacuum
rest frame") can exist without contradicting the minkowskian structure of the
space-time felt by "ordinary" particles (those with critical 
speed equal to $c$).
Then, $c$ will not necessarily be the only critical speed in vacuum: for
instance, superluminal sectors of matter may exist related to new degrees
of freedom not yet discovered experimentally. Such particles would not be
tachyons: they may feel different minkowskian space-times with critical speeds
$c_i\gg c$ (the subscript $i$ stands for the $i$-th superluminal sector),
and behave kinematically like "ordinary" particles apart from the difference in
critical speed.
A superluminal sector of matter can be built as follows.

Ordinary free particles in vacuum usually satisfy a dalembertian equation, such
as the Klein-Gordon equation for scalar particles:
\equation
(c^{-2}~\partial ^2 /\partial t^2~-~\Delta)~\phi~+~m^2c^2~(h/2\pi
)^{-2}\phi~=~0
\endequation
where the coefficient of the second time derivative sets $c$ , the critical
speed in vacuum (speed of light). Given $c$ and the Planck
constant $h$ , the coefficient of the linear term in $\phi $ sets $m$ , the
mass
. To study solutions of the wave equation, we consider the
conserved observables:
\equation
E~=~i~(h/2\pi )~\partial /\partial t ~~~~~~~~~~,~~~~~~~~~~\vec {\mathbf
p}~=~-i~(h/2\pi )~\vec {\nabla }
\endequation
and get plane wave solutions from which we can build position and speed
operators \cite{sch} . In the non-relativistic limit, it can be checked that 
$m$ is indeed the inertial mass \cite{gon1} .
With the conservative choice of leaving the Planck constant unchanged,
superluminal sectors of matter can be generated replacing in
the above construction the speed of light $c$ by a new critical speed $c_i$
for
the $i$-th superluminal sector. All previous concepts remain
valid, leading to particles with positive mass and energy which are not
tachyons. For inertial mass $m$ and critical speed $c_i$ , the new particles
will have rest energies \cite{gon1} :
\equation
E_{rest}~=~mc_i^2
\endequation
which, for a given inertial mass, are much higher than the rest energies of
"ordinary" particles. This generalization of the Einstein equation implies
in particular that: a) in accelerator experiments, very high energies can be
required to produce superluminal particles; b) cosmic ray events
originating from superluminal particles can release very high energies.

\subsection{A scenario with several critical speeds in vacuum}

In what follows, we shall consider a scenario with several sectors of matter:
a) the "ordinary" sector, made of "ordinary" particles with a critical speed
equal to the speed of light $c$ ; b) one or more superluminal sectors, where
particles have critical speeds $c_i \gg c$ in vacuum, and each sector is
assumed to have its own Lorentz invariance with $c_i$ defining the metric.

If the standard minkowskian space-time is not a compulsory framework, we can
conceive fundamentally different descriptions of space and time \cite{gon2} . 
Space-time
can, for instance, be galilean, or minkowskian with an absolute critical speed 
$C \gg c$ . Another
possibility, requiring an absolute origin, 
would be to consider a $SU(2)$ spinorial space-time where time 
would correspond to the spinor modulus (a $SU(2)$ scalar, 
positive definite and therefore
setting an arrow of time), and the three space dimensions would originate
from the tangent 
hyperplane to the $S^3$ hypersphere of constant modulus in the
$C^2$ (topologically equivalent to $R^4$) spinor space. In this tangent 
hyperplane,
the three independent 
directions correspond to the three $SU(2)$ generators and therefore
define a vector representation of $SU(2)$ .

Even if each sector has its own "Lorentz invariance" involving as the basic
parameter the critical speed in vacuum of its own particles, interaction
between two different sectors will break both Lorentz invariances. 
The concept of mass, as a relativistic invariant, will become approximate and
sectorial. In our approach,
the vacuum is a material medium as suggested by recent results in particle
physics, and the Michelson-Morley result is not incompatible with the existence
of some "ether" defining an absolute local 
rest frame (the "vacuum rest frame"). 
If superluminal particles couple weakly to ordinary matter,
their effect on the ordinary sector will occur at very high energy and short
distance, far from the domain of successful conventional tests of Lorentz
invariance. 
The actual structure of space and time will be found only by going 
beyond the above wave equations to deeper levels of resolution, similar to the
way high-energy accelerators explore the inner structure of "elementary"
particles. 

Our scenario is far from being the first case in which several critical speeds
coexist in a medium. In a perfectly transparent crystal close to zero
temperature, two critical speeds exist: the speed
of sound and the speed of light.

\section{Dynamics and cosmology}

Mass mixing between particles from different sectors may occur and,
although very weak, be more significant for very light 
particles (e.g. photons,
neutrinos...). Since the graviton is an "ordinary" gauge boson, associated to
ordinary Lorentz invariance, it is not expected to play a universal role in
the presence of superluminal particles. Assuming that each superluminal sector
has its own Lorentz metric $g_{[i]\mu \nu }$ ($[i]$ for the $i$-th
sector), with $c_i$ setting the speed scale, we may expect each sector to
generate its own gravity with a coupling constant $\kappa _i $ and a
sectorial graviton traveling at speed $c_i$ . "Gravitational" interactions
between different sectors will be weak and concepts so far considered
as very fundamental (e.g. the universality of the exact equivalence between
inertial and gravitational mass) will become approximate sectorial properties.

If superluminal sectors couple to ordinary matter, they are expected to release 
"Cherenkov" radiation (e.g. spontaneous emission of particles whose critical
speed is lower than the speed of the particle) in vacuum when they move at
a speed $v > c$ . Thus, superluminal particles will be eventually decelerated
to a speed $v \leq c$ . The nature and rate of "Cherenkov" radiation in vacuum
will depend on the superluminal particle and can be very weak in some cases.
In accelerator experiments, this "Cherenkov" radiation may provide a clean
signature allowing to identify some of the produced superluminal particles.

If each sectorial Lorentz invariance is expected to break down below a critical
distance scale $k_i^{-1}$ , $k_o^{-1}$ for the ordinary sector, where the
$k_i$ and $k_o$ are critical wave vector scales, we can expect 
\cite{gon3} the appearance
of critical temperatures $T_o$ and $T_i$ defined by:
\equation
kT_o~\approx~\hbar c k_o~~~~~~~~~~,~~~~~~~~~~kT_i~\approx~\hbar c_i k_i
\endequation
defining phase transitions in field theories, as well as in the very early
Universe. These singularities seem to prevent conventional extrapolations to a
Big Bang limit. Above $T_o$ , the Universe may have contained only superluminal
particles and dynamical correlations have been able to propagate much faster
than light. This invalidates standard arguments leading to the so-called
"horizon problem" and "monopole problem". Conventional Friedmann equations will
not hold in the new scenario, and the need for inflation is far from obvious.
In the above considered spinorial space-time, the Big Bang limit can possibly be
related to the absolute origin in the spinor space. In this approach, it seems
impossible to set a "natural time scale" based on extrapolations (e.g. to
Planck time) from our
knowledge of the low energy sector. Arguments leading to the "flatness" and
"naturalness" problem, as well as the concept of the cosmological constant and
the relation between critical density and Hubble's "constant" (one of the basic
arguments for ordinary dark matter at cosmic scale) should be reconsidered.
Superluminal particles may have played a cosmological role leading to
substantial changes in the "Big Bang" theory. They can be very abundant and
even provide nowadays most of the (dark) matter at cosmic scale, therefore
leading the present evolution of the Universe.
 
\section{Superluminal particles and dark matter identification}

If superluminal particles are very abundant, they can, in spite of their
expected weak coupling to "ordinary" gravitation, produce some observable
gravitational effects. It is not obvious how to identify the superluminal
origin of a collective gravitational phenomenon, but signatures may exist, 
e.g. in gravi- tational collapses or if it were possible to detect
superluminal gravitational waves. We do not in general expect
concentrations of superluminal matter to follow those of ordinary matter, but
this is not excluded, e.g. in the presence of coupled gravitational
singularities involving several sectors. 
If astrophysical concentrations of
superluminal particles produce high-energy particles, cosmic rays may provide a
unique way to detect such objects \cite{gon4}. Direct detection of particles
from superluminal matter around us, e.g. in underground and underwater
detectors, should not be discarded \cite{gon2} . At very high energy, they
can escape the Greisen-Zatsepin-Kuzmin cutoff \cite{GZK} and be at
the origin of the highest-energy events. At lower energies, they can
produce detectable signals.

\subsection {Superluminal primaries}
 
High-energy superluminal particles can be produced from acceleration, decays,
explosions... in astrophysical objects made of superluminal matter, or from 
"Cherenkov" emission in vacuum by particles with higher critical speed.
They can reach the earth and undergo
collisions inside the atmosphere, producing many secondaries like ordinary
cosmic rays. They can interact with the rock or with water near some
underground or underwater detector, coming from the atmosphere or after having
crossed the earth. Contrary to neutrinos, whose flux is attenuated
by the earth at energies above 10${^6}~ GeV$ , superluminal particles will in
principle not be stopped by earth at these energies.
Such primaries can release most of their energy in inelastic collisions, and 
rather high energies (with momentum transfer of the order of the incoming 
momentum) in elastic 
scattering.
Low-energy superluminal particles can also produce detectable events.
At $v\simeq c$ (after "Cherenkov" deceleration in vacuum),
superluminal primaries can produce recoil protons and neutrons in the GeV range
and inelastic events of higher energies. Such events would be detectable,
e.g. in Cherenkov detectors for neutrino astronomy, even at very small rates.
In cryogenic detectors, unconventional recoil spectra (e.g. indicating an
escape velocity much above 10$^{-3}c$) can be a signature for
superluminal dark matter.

\subsection{Ordinary primaries}

Annihilation of pairs of slow superluminal particles, releasing very high
kinetic energies from the relation $E=mc_i^2$ , can be a source of high-energy
ordinary and superluminal cosmic rays. Decays and "Cherenkov" radiation in
vacuum can produce similar effects. Thus, ordinary cosmic rays can be produced
anywhere and not just near astrophysical objects made of ordinary matter.


\section*{References}

\begin{thebibliography}{99}

\bibitem{sch} See, e.g. S.S. Schweber, "An Introduction to Relativistic Quantum Field
Theory", Row, Peterson and Co. 1961 .
\bibitem{gon1}L. Gonzalez-Mestres, "Properties of a possible class of particles
able to travel faster than light", Proceedings of the Moriond Workshop on
"Dark Matter in Cosmology, Clocks and Tests of Fundamental Laws", Villars
(Switzerland), January 21-28 1995 , Ed. Fronti\`eres, France. Paper
astro-ph/9505117 of electronic library.
\bibitem{gon2}L. Gonzalez-Mestres, "Physical and cosmological implications of a
possible class of particles able to travel faster than light", Contributed
paper to the 28$^{th}$ International Conference on High Energy Physics, Warsaw
(Poland), July 25-31 1996 . 
\bibitem{gon3}L. Gonzalez-Mestres, "Cosmological implications of a possible
class of particles able to travel faster than light", Proceedings of the IV
International Conference on Theoretical and Experimental Aspects of Underground
Physics, TAUP 95 , Toledo (Spain), September 1995 , Ed. Nuclear Physics
Proceedings.
\bibitem{gon4}L. Gonzalez-Mestres, "Superluminal particles and high-energy
cosmic rays", May 1996 . Paper astro-ph/9606054 of electronic library.
\bibitem{GZK}K. Greisen, Phys. Rev. Lett. 16, 748 (1966); G.T. Zatsepin and
V.A. Kuzmin, Pisma Zh. Eksp. Teor. Fiz. 4 , 114 (1966).



\end{thebibliography}
\end{document}